\documentclass[12pt]{iopart}

\usepackage{latexsym,amssymb}
\usepackage{graphics}
\usepackage{graphicx}
\usepackage{epsfig}

\renewcommand{\phi}{\ensuremath{\varphi}}

\begin{document}

\title{Factorised Steady States in
Mass Transport Models}
\author{M.\ R.\ Evans$^1$, Satya\  N.\ Majumdar$^{2,3}$, R.\ K.\ P.\ 
Zia$^{4,5}$}

\address{
$^1$School of Physics, University of Edinburgh,\\
Mayfield Road, Edinburgh EH9 3JZ, UK\\[0.5ex]
$^2$ Laboratoire de Physique Theorique et Modeles
Statistiques,\\ Universite Paris-Sud, Bat 100, 91405, Orsay-Cedex,
France\\[0,5ex]
$^3$Laboratoire de Physique Theorique (UMR C5152
du CNRS),\\ Universite Paul Sabatier, 31062 Toulouse Cedex, France\\[0.5ex]
$^4$Department of Physics and\\
Center for Stochastic Processes in Science and Engineering,\\
Virginia Tech, Blacksburg, VA 24061-0435, USA;\\[0.5ex]
$^5$FB-Physik, Universit\"at Duisburg-Essen, 45117 Essen, Germany.}

\date{April 26, 2004}

\begin{abstract}
  We study a class of mass transport models where mass is transported
  in a preferred direction around a one-dimensional periodic lattice
  and is globally conserved. The model encompasses both discrete and
  continuous masses and parallel and random sequential dynamics and
  includes models such as the Zero-range process and Asymmetric random
  average process as special cases. We derive a necessary and
  sufficient condition for the steady state to factorise, which takes
  a rather simple form.

\end{abstract}

\pacs{05.70.Fh, 02.50.Ey, 64.60.-i}

%\submitto{\JPA}

\maketitle

%**************************************************************************%
%********************************* main text ******************************%
%**************************************************************************%
Mass transport models form a general class of lattice models defined
by dynamics in which mass is transferred (without loss) stochastically
from site to site. They have attracted much recent attention,
especially in connection with ``condensation transitions''
\cite{MRE00,E96,BBJ,OEC,MKB,LMZ}. Examples include the Zero-Range Process
(ZRP) \cite{MRE00} and Asymmetric Random Average Process (ARAP)
\cite{KG,RM1}, which have been used to model such diverse situations
as traffic flow, clustering of buses {\cite{OEC}, phase separation
dynamics \cite{KLMST} and force propagation through granular media
\cite{CLMNW}. In general, it is difficult to determine the steady
state distribution of such models. Thus, it is remarkable that, not
only are the steady states of many models found, they often share a
very convenient property, namely, a factorised steady state (also referred
to as a product measure). Of course, such a property greatly
facilitates the analysis of interesting behaviour, e.g.,
condensation. }

In this letter we determine the requirement for a factorised steady state in
a very broad class of mass transport models. The form of this necessary and
sufficient condition, stated in (\ref{answer!}), turns out to be appealingly
simple. Encompassing both random sequential and parallel dynamics, this
class includes both the ZRP and ARAP. 
We  discuss the salient features of the approach leading to (\ref{answer!})
and  recover some previously known cases.

We consider a one-dimensional lattice of $L$ sites with periodic boundary
conditions (site $L+1$= site 1): associated with each site is a mass $m_i$, $%
i=1\ldots L$. The total mass is given by $M=\sum_{i=1}^Lm_i$. We shall most
generally consider $m_i$ as continuous variables. The dynamics is as
follows: from time $t$ to $t+1$, at each site $i$, mass $\mu _i$ drawn from a
distribution $\phi (\mu _i|m_i)$ `chips off' the mass $m_i$, and moves to
site $i+1$. Thus the master equation for the weights (unnormalised
probabilities) $F_t(\underline{m})$ is 
\begin{equation}
F_{t+1}(\underline{m})=\prod_{i=1}^L\int_0^\infty \ensuremath{\mathrm{d}}%
m_i^{\prime }\int_0^{m_i^{\prime }}\ensuremath{\mathrm{d}}\mu _i\ \phi (\mu
_i|m_i^{\prime })\prod_{j=1}^L\delta (m_j-m_j^{\prime }+\mu _j-\mu
_{j-1})\,F_t(\underline{m^{\prime }}) \;, 
\label{Meq}
\end{equation}
where $\underline{m}\equiv \{m_1,m_2,\dots ,m_L\}$. 
Note that this dynamics conserves the total mass, $M$, so that $F_t(%
\underline{m})$ may be considered as a function of only $L-1$ variables.
%Defined on a \emph{compact} subspace, 
The integral of the weights, subject to the constraint
of globally conserved mass,
\begin{equation}
Z\left( M,L\right) \equiv \prod_{i=1}^L\int_0^\infty \ensuremath{\mathrm{d}}%
m_i\, \delta \left( M-\sum_{i=1}^Lm_i\right) F_t(\underline{m})\,\,,  \label{Z}
\end{equation}
should be finite and serves as a ``partition function,'' so that $F/Z$ is a
probability density (or ``canonical distribution'').

In the $t\to \infty$ limit, $F_t(\underline{m})$ approaches a 
$t$-independent function, which we denote simply by $F(\underline{m})$ 
and refer to as the steady state.
%We assume that the steady state is unique.
 Defining the Laplace transform 
\begin{equation}
G(\underline{s})= \left[ \prod_{i=1}^L \int_0^\infty \ensuremath{\mathrm{d}}
m_i e^{-s_i m_i} \right] F(\underline{m}) \;,
\end{equation}
and transforming (\ref{Meq}), we find 
\begin{eqnarray}
G(\underline{s}) &=& \left[ \prod_{i=1}^L \int_0^\infty \ensuremath{%
\mathrm{d}} m_i^{\prime}\int_0^{m_i^{\prime}} \ensuremath{\mathrm{d}} \mu_i\
\phi(\mu_i |m_i^{\prime}) e^{-s_i\left( m_i^{\prime }- \mu_i
+\mu_{i-1}\right) } \right] F(\underline{m}^{\prime})\;.  \label{LT}
\end{eqnarray}

We now assume that the steady state weight factorises 
\begin{equation}
F(\underline{m})=\prod_i f(m_i)  \label{Ffac}
\end{equation}
which implies 
\begin{equation}
G(\underline{s})=\prod_ig(s_i)\quad \mbox{where}\quad g(s)=\int_0^\infty
dmf(m)e^{-sm}\;.
\end{equation}
Then (\ref{LT}) becomes 
\begin{equation}
\prod_i g(s_i)=\prod_i\left[ \int_0^\infty dm_i^{\prime }f(m_i^{\prime})
\int_0^{m_i^{\prime}}\ensuremath{\mathrm{d}} \mu_i\,\phi (\mu_i|m_i^{\prime
}) e^{-s_i\left( m_i^{\prime }- \mu_i +\mu_{i-1}\right) } \right]\;.
\label{cond1}
\end{equation}
Changing variables to $\sigma \equiv m-\mu$ (the mass 
remaining after the move), we write 
\begin{equation}
f(m) \phi (\mu|m ) = \mathcal{P} (\mu,\sigma)\;.  \label{phimun}
\end{equation}
Note that no assumption on the form of $f(m)$ or $\phi (\mu|m )$ is implied
at this point. With this notation (\ref{cond1}) becomes 
\begin{equation}
\prod_ig(s_i)=\prod_i\left[ \int_0^\infty \ensuremath{\mathrm{d}}%
\mu_i\,\int_0^\infty \ensuremath{\mathrm{d}}\sigma_i\, \mathcal{P}%
(\mu_i,\sigma_i) e^{-s_i\sigma_i - s_{i+1}\mu_i} \right]\; .  \label{fac}
\end{equation}

A necessary and sufficient condition for the solution of (\ref{fac}), is 
\begin{equation}
\int_0^\infty \ensuremath{\mathrm{d}}\mu_i\,\int_0^\infty %
\ensuremath{\mathrm{d}}\sigma_i\, \mathcal{P}(\mu_i,\sigma_i)
e^{-s_i\sigma_i - s_{i+1}\mu_i} = \ell(s_i) k(s_{i+1})  \label{cond}
\end{equation}
where the two functions, $k$ and $\ell $, must satisfy 
\begin{equation}
k\left( s\right) \ell \left( s\right) =g\left( s\right) .  \label{kl}
\end{equation}
That (\ref{cond}) is necessary and sufficient may be seen by taking the
logarithm of (\ref{fac}) then taking derivatives with respect to $s_i$ and $%
s_{i+1}$.

Condition ({\ref{kl}) implies via the convolution theorem that 
\begin{eqnarray}
f\left( m\right) =\left[ v*w\right] \left( m\right) \equiv \int_0^md\mu \
v\left( \mu \right) w\left( m-\mu \right)   \label{fm}
\end{eqnarray}
where 
\begin{equation}
k\left( s\right) =\int_0^\infty d\mu \,e^{-s\mu }v\left( \mu \right) \quad
;\quad \ell \left( s\right) =\int_0^\infty d\sigma \,e^{-s\sigma }w\left(
\sigma \right) .  \label{kldef}
\end{equation}
Then, equations (\ref{cond}) and (\ref{kldef}) imply 
\begin{equation}
\mathcal{P}\left( \mu ,\sigma \right) =v\left( \mu \right) w\left( \sigma
\right) \;.  \label{P}
\end{equation}
Finally we obtain from (\ref{phimun}) and (\ref{fm}) 
\begin{equation}
\phi (\mu |m)=\frac{v\left( \mu \right) w\left( m-\mu \right) }{\left[
v*w\right] \left( m\right) }.  \label{answer!}
\end{equation}
Let us emphasize that the condition for a factorised stationary distribution
for the whole lattice precisely reduces to the condition that $\phi (\mu |m)$
has the form (\ref{answer!}). Thus, equation (\ref{answer!}) is the central
result of this paper: for chipping rules of the form (\ref{answer!}), one has
a factorised steady state (\ref{Ffac}) with weights given by (\ref{fm}). Let
us comment on several important points. Equations (\ref{kl}-\ref{answer!})
allow us to define ``equivalence classes'' of chipping distributions---those
leading to the same stationary state---by dividing }$k\left(
s\right) $ and multiplying $\ell \left( s\right) ${\ by any (well behaved)
function of }$s$. In particular, their roles can be ``reversed'' to form a
``dual'' $\phi $, i.e., $w\left( \mu \right) v\left( m-\mu \right) /\left[
w*v\right] \left( m\right) $. Now, we can obviously interpret the factors in
(\ref{P}) as a function for $\mu $, the mass which moves, and a function of $%
\sigma $, the mass which stays. In this sense, ``duality'' reverses these
two portions of the mass, without changing $F(\underline{m})$. If we further
perform a Galilean transformation (shifting the entire lattice by one site
in a time step) and a parity transformation ($i\Leftrightarrow L+1-i$), then
we recover the original system.
Finally, note that both $\phi $ and the steady state ($F/Z$) are
invariant under shifts of $\ln v$ and $\ln w$ by a linear function (i.e.,
there are arbitrary amplitudes or exponential factors
in $v$ and $w$: $a^\mu , a^\sigma $).

In addition to treating models with parallel dynamics, manifest in (\ref
{Meq}), we can extend the approach outlined above
to models with random sequential
dynamics. Let the probability of a chipping event in a time step $\propto dt$
so that, to leading order in $dt$, at most one chipping event over the whole
lattice occurs per update. Furthermore, we can let the duration of a time
step be $dt$ and take $dt\to 0$ to obtain a continuous time limit where
chipping events occur with rates per unit time. Thus, we write 
\begin{equation}
v(\mu )=\delta (\mu )+x(\mu )dt,
\end{equation}
where $\delta (\mu )$ is the Dirac delta function. Then (\ref{fm}) and (\ref
{answer!}) yield 
\begin{eqnarray*}
f(m) &=&w(m)+dt[x*w](m) \quad \mbox{and} \\[1ex]
\phi (\mu |m) &=&\frac 1{w(m)+dt[x*w](m)}\left\{ \delta (\mu
)w(m)+dt\,x(\mu )w(m-\mu )\right\} \\[1ex]
&=&\delta (\mu )\left[ 1-\frac{dt}{w(m)}\left[ x*w\right] (m)\right] +dt%
\frac{x(\mu )w(m-\mu )}{w(m)}+O(dt^2)\;.
\end{eqnarray*}
Taking $dt\to 0$ we obtain the continuous time limit where mass $\mu $ moves
from a site with mass $m$ with rate $x(\mu )w(m-\mu )/w(m)$ and $f(m)=w(m)$.

Let us illustrate how this approach unifies two seemingly unrelated models
 -- ARAP and ZRP.
First we consider the ARAP \cite{KG,RM1,ZS1,ZS2}, a model in which 
each site contains a continuous amount of mass and at each time step a 
random fraction of the mass moves to the next site to the right. Its precise 
definition lies in $\psi (r|m)=\phi (\mu |m)m$, the distribution for $r$, 
the fraction of mass that moves to the neighbouring site. A known family of 
distributions where one has a factorised steady state is 
$\psi (r|m)=(n-1)r^{n-2}$ \cite{CLMNW,ZS1} which becomes 
\begin{equation}
\phi (\mu |m)=(n-1)\frac{\mu ^{n-2}}{m^{n-1}}\;.  \label{sARAP}
\end{equation}
In our approach, the results are particularly simple: 
\begin{eqnarray}
v(\mu ) &=&\mu ^{n-2},\,\,\,w(\sigma )=1,  \label{ARAPvw} \\
f(m) &=&m^{n-1}/\left( n-1\right) \,\,.  \label{ARAPf}
\end{eqnarray}
Note that, to relate this $f\left( m\right) $ to relevant quantities in the
literature (e.g., \cite{ZS1}), the single site mass distribution, defined as
the full distribution integrated over the rest of the mass variables, is $%
p\left( m\right) =f\left( m\right) Z\left( M-m,L-1\right) /Z\left(
M,L\right) $. In this case, $Z\left( M,L\right) =M^{nL-1}\left[ \Gamma
(n-1)\right] ^L/\Gamma (nL)$, so that our $p\left( m\right) 
$ reduces, e.g., to equation (37) of \cite{ZS1} in the thermodynamic limit.

Another well-known case is the Zero-Range Process, reviewed in \cite{MRE00}.
A focus of major interest (for recent developments see for example \cite
{EH03,GSS,Godreche,AEM}), it is a mass transport model where $m_i$ takes integervalues and a unit mass moves from site $i$ to site $i+1$ with probability $%
u(m_i)$. 
Within our approach, this model appears as a very special case,
with Dirac delta distributions for both $v$ and $w$. Since the moved mass
can take only two values while the one remaining can be of any integer, the
most general forms are 
\begin{eqnarray}
v(\mu )=\delta (\mu )+a\delta (\mu -1),\quad w(\sigma )=\sum_{k=0}^\infty
w_k\delta (\sigma -k)\;,  \label{zrpVW}
\end{eqnarray}
where $a$ and $w_k$ are arbitrary weights. As overall amplitudes are
irrelevant, we have chosen the coefficient of $\delta \left(\mu \right) $
to be unity and will set $w_0=1$. From $f=v*w$, we see that
\begin{eqnarray}
f(m) &=& w(m)+aw(m-1)\\
%\delta (m)+\sum_{k=1}^\infty \left[ w_k\delta (m-k)+aw_k\delta
%(m+1-k)\right]  \\
&=&\delta (m)+\sum_{k=1}^\infty \left[ aw_{m-1}+w_m\right] \delta (m-k)\;.
\label{zrpss}
\end{eqnarray}
With a little care, we obtain 
\begin{equation}
\phi (\mu |m)=\frac{w_m\delta (\mu )+aw_{m-1}\delta (\mu -1)}{w_m+aw_{m-1}}.
\end{equation}
The coefficient of $\delta (\mu -1)$ is precisely the chipping probability,
denoted by $u\left( m\right) $ above. From here, we easily find the $w_m$
in terms of the $u$: 
\begin{equation}
w_m=a^m\prod_{n=1}^m\frac{1-u(n)}{u(n)}\;.
\end{equation}
Substituting this expression into (\ref{zrpss}) yields
for  the weights, 
\begin{equation}
f(m)=\sum_{k=0}^\infty
f_k\delta (m-k)
\end{equation}
where
\begin{eqnarray}
f_k &=& \frac{a^k}{1-u(k)}\prod_{n=1}^k\frac{1-u(n)}{u(n)}
\quad \mbox{for}\quad k \geq1\;,
 \label{zrpss-MRE}
\\
&=& 1 
\quad \mbox{for}\quad k =0 \;.
\end{eqnarray}
This result was previously obtained by a more complicated approach \cite{MRE97}.
Note that the factors $a^k$ will drop out when we consider the probability
density itself: $F(\underline{m})/Z$. We close this paragraph by noting the
case with random sequential dynamics, which is obtained by letting $a=\tilde{%
a}\ensuremath{\mathrm{d}}t$ and $u(m)=x(m)\ensuremath{\mathrm{d}}t$ where $%
\ensuremath{\mathrm{d}}t\to 0$ yielding 
\begin{equation}
f_k=\tilde{a}^k\prod_{n=1}^k\frac 1{x(k)}\quad k \geq 1\;.
\end{equation}

Finally, the results presented here may be generalised to the case of
heterogeneous mass transfer where $\phi _i(\mu |m)$ depends on the site $i$.
A necessary and sufficient condition for a factorised steady state is that 
\begin{equation}
\phi _i(\mu |m)=\frac{v\left( \mu \right) w_i\left( m-\mu \right) }{\left[
v*w_i\right] \left( m\right) }
\end{equation}
where $v$ and $w_i$ are arbitrary functions but $v$ must be the same for
each site. The weight functions are given by 
\begin{equation}
f_i(m)=\left[ v*w_i\right] \left( m\right) \,\,.
\end{equation}

To conclude, we have determined the condition for the steady state in a
general class of mass transport models to factorise. This class encompasses
both continuous and discrete mass, as well as parallel and random sequential
dynamics. Not only does this approach provide a unified perspective of all
previously known models, it opens avenues to construct new models with this
property (e.g., binomial chipping process and generalized Zero-Range
Processes). In addition, we believe this approach would facilitate a deeper
understanding of the existence and nature of condensates and possibly reveal
novel forms of phase transitions. Implications of the gauge-like 
transformations should also be explored. Further work is in progress and will be
published elsewhere.

\ack

We thank Max Planck Institute, Dresden where this work was initiated for
hospitality. One of us (RKPZ) thanks H.W.~Diehl for his hospitality at the
University of Duisburg-Essen, where some of this research was carried out,
and acknowledges the support from the Alexander von Humboldt Foundation and
the US National Science Foundation through DMR-0088451. 

\end{document}